\documentclass{osa-article}

\journal{oe}

\articletype{Research Article}

\usepackage{lineno}

\begin{document}

\title{Non-invasive color imaging through scattering medium under broadband illumination}

\author{Yunong Sun,\authormark{1} Jianbin Liu,\authormark{1,*} Hui Chen,\authormark{1} Zhuoran Xi,\authormark{1} Yu Zhou,\authormark{2} Yuchen He,\authormark{1} Huaibin Zheng,\authormark{1} Zhuo Xu,\authormark{1} Yuan Yuan,\authormark{1,*} }

\address{\authormark{1}Electronic Materials Research Laboratory, Key Laboratory of the Ministry of Education \& International Center for Dielectric Research, Xi'an Jiaotong University, Xi'an 710049, China}

\address{\authormark{2}MOE Key Laboratory for Nonequilibrium Synthesis and Modulation of Condensed Matter, Department of Applied Physics, Xi’an Jiaotong University, Xi’an, Shaanxi 710049, China}

\email{\authormark{*}liujianbin@mail.xjtu.edu.cn}



\begin{abstract}
Due to the complex of mixed spectral point spread function within memory effect range, it is unreliable and slow to use speckle correlation technology for non-invasive imaging through scattering medium under broadband illumination. The contrast of the speckles will drastically drop as the light source's spectrum width increases. Here, we propose a method for producing the optical transfer function with several speckle frames within memory effect range to image under broadband illumination. The method can be applied to image amplitude and color objects under white LED illumination. Compared to other approaches of imaging under broadband illumination, such as deep learning and modified phase retrieval, our method can provide more stable results with faster convergence speed, which can be applied in high speed scattering imaging under natural light illumination.
\end{abstract}

\section{Introduction}
How to penetrate random scattering media for quick and reliable non-invasive imaging has become one of the hottest issues in the field of computational imaging, with applications ranging from biological imaging to astronomical imaging\cite{Speckle:Phenomena,RN35,RN85,RN89}. Different methods for non-invasive scattering imaging have been developed recently. Wavefront shaping can be used to image objects behind scattering medium by controlling the spatial light modulator\cite{RN51,RN52,RN122}. However, this method usually demands reference objects like a guide star in the plane of interest. Measuring the transmission matrix of the entire scattering system is also an non-invasive imaging method\cite{RN123,RN124}. The determination of the transmission matrix takes long time and requires high precision. Alternatively, it is relatively efficient to measure the point spread function (PSF) of the scattering system and perform a deconvolution operation\cite{RN65,RN98}. However, the method needs prior information and will lose the important advantage of non-invasion imaging.

A breakthrough in the field of non-invasive imaging is on speckle correlation technology (SCT) within optical memory effect range and phase retrieval algorithms\cite{RN31}. The speckle correlation method based on memory effect draws conclusions from autocorrelation of PSF of the scattering system. Within memory effect range, the speckle autocorrelation approximates the autocorrelation of the object, as shown in Eq. (\ref{eq:autocorrelation})\cite{RN31}, .
\begin{equation}
\begin{aligned}
I\star I=(O\star O)\ast (S\star S) \approx O\star O, \label{eq:autocorrelation}
\end{aligned}
\end{equation}
where $I$ is the speckle captured by camera, $S$ is the PSF of the imaging system and $O$ is the object in space domain. $\star$ denotes the autocorrelation operation, and $\ast$ denotes the convolution operation.

The key to Eq. (\ref{eq:autocorrelation}) is $S\star S$, which is a sharp peak function. The autocorrelation of PSF can be ignored in the convolution operation\cite{Li:19,RN91}. So strict requirement of PSF limits SCT-based method valid for narrow band illumination. Deep learning is an existing reliable method for imaging through scattering media under broad-spectrum illumination\cite{RN40,RN101}. However the method requires a large amount of sample data for an end-to-end learning. Another method is to modify the phase retrieval algorithm by introducing constraints of the phase of OTF (PhTF)\cite{RN91}. However the iterative process of the phase retrieval algorithm is unstable and the calculation process can take a long time\cite{Fienup:82}.

Broadband spectrum and multi-spectrum are an essential part in color imaging, in which images from different spectrum are superposed to recover the color or the spectral information of object\cite{RN65}. The current widely used method is to measure the spectral PSF of several discrete spectrum and apply deconvolution or correlation operations on each of them\cite{RN65,RN98}. By this way, the object can be rebuilt with deconvolution operation as Eq. (\ref{eq:deconv.})\cite{MORRIS1997197},

\begin{equation}
\begin{aligned}
O=\sum_{\lambda}O_{\lambda}
=\sum_{\lambda}\mathscr{F}^{-1}(\mathscr{F}(O_{\lambda}))
=\sum_{\lambda} \mathscr{F}^{-1}(\frac{\mathscr{F}(I_{\lambda})}{e^{-i\Phi_{s_{\lambda}}}}),\label{eq:deconv.}
\end{aligned}
\end{equation}
where $I$ is the speckle captured by camera, $\Phi_S$ is the PhTF of the imaging system, $\lambda$ is the wavelength of light source, $O$ is the object in space domain, $\mathscr{F}$ and $\mathscr{F}^{-1}$ denotes the Fourier transform and inverse Fourier transform, respectively.

However, manipulating spectral PSF can obtain the orientation and position information additionally. Those approaches are limited by the need for prior information of spectral PSF and can not be applied in non-invasion circumstance. As an alternative, a non-invasive triple correlation-based color image reconstruction method was suggested \cite{RN106}. It contains orientation information which is missing during the traditional speckle correlation. But triple correlation algorithm requires narrow band illumniation source as SCT and it also takes long time to complete than SCT with phase retrieval process\cite{RN37}.

In this paper, The multi-frame OTF retrieval engine (\emph{MORE})\cite{Chen:2020} is used to achieve non-invasive color imaging under broadband illumination by generating OTF with speckles from different objects. It can not only breaks the limitation of speckle correlation imaging technology, but also have stable and fast iterative convergence due to redundant information of speckles.
\section{Principle and methods}
\label{sec:System setup and analytical results}

\begin{figure}[htbp]
\centering
\includegraphics[width=12cm]{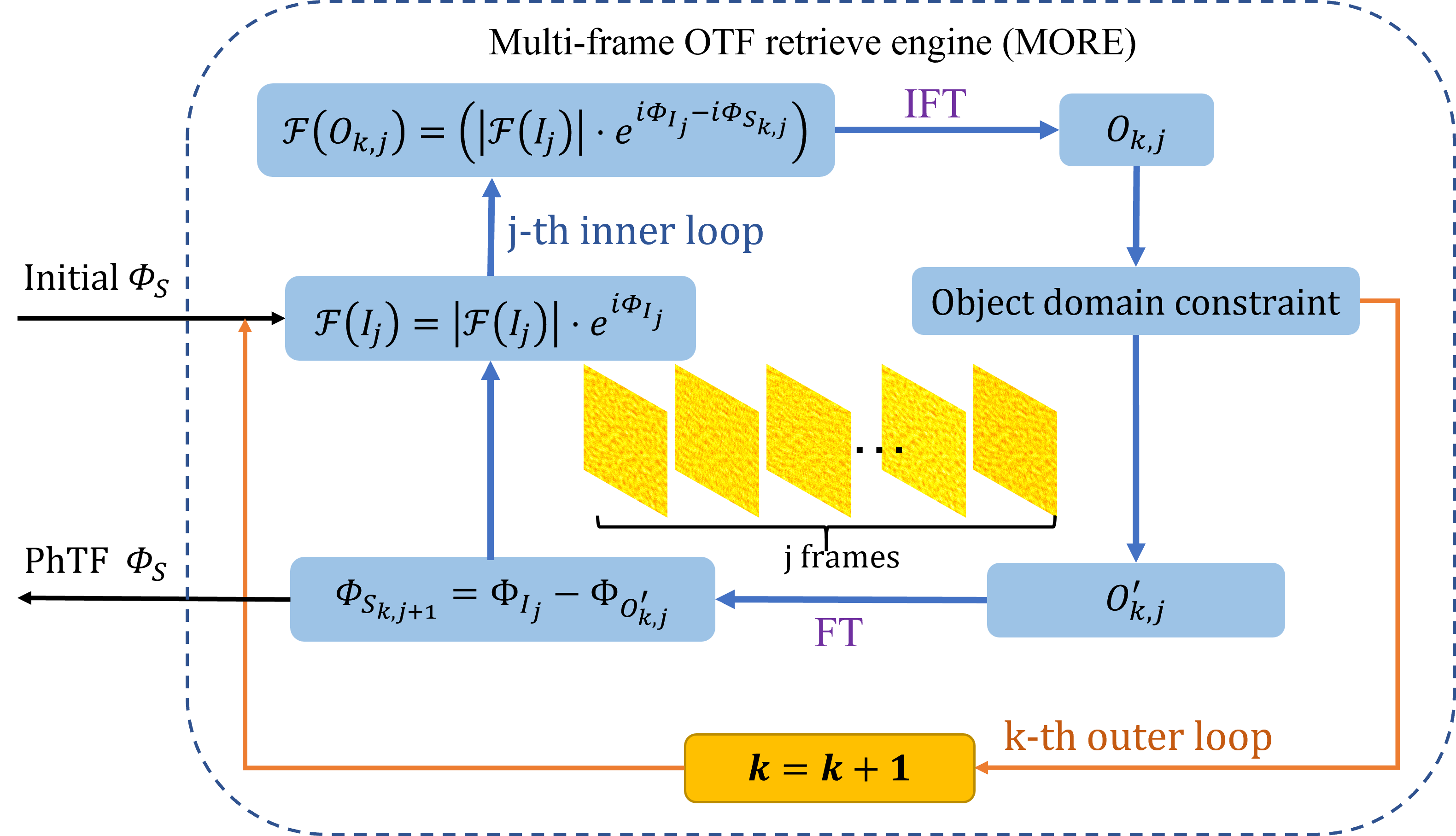}
\caption{\label{fig:MORE_block}Block diagram of the \emph{MORE} for dynamic imaging object hidden behind the scattering medium under broadband illumination. The retrieval process starts with an inital guess, $\Phi_{S}$. Inner loop is a iteration process similar to Error-Reduction algorithm. The outer loop repeats the iterative process of the inner loop until the set number of times is reached. Then the updated PhTF, $\Phi_{S}$, is obtained. Where $|\mathscr{F}(I_j)|$ denotes the amplitude of Fourier transform of j-th frames and $\Phi_{S_{k,j+1}}$ indicates the updated PhTF at the k-th outer loop and j-th inner loop.}
\end{figure}

\emph{MORE}, has successfully carried out high-speed dynamic imaging under low signal-to-noise ratio circumstances in our earlier work\cite{Chen:2020}. When the PSF and the speckle for the system's object have the relationship $I=O\ast S$, the relationship can be expressed in the frequency domain as follows:

\begin{equation}
\begin{aligned}
\mathscr{F}(I)=\mathscr{F}(O) \cdot \mathscr{F}(S), \label{eq:I=O*S}
\end{aligned}
\end{equation}
where $I$ is the speckle pattern, $O$ is the object, and $S$ for the PSF of the system.

Rewrite Eq. (\ref{eq:I=O*S}) as
\begin{equation}
\begin{aligned}
\mathscr{F}(I) \cdot e^{-i\Phi_S}=|\mathscr{F}(S)|\cdot |\mathscr{F}(O)|e^{i\Phi_{O}},\label{eq:fourier(I=O*S)}
\end{aligned}
\end{equation}
in which $|\mathscr{F}(S)|$ is the amplitude of the OTF. It is proved that amplitude of the OTF only acts as a spatial frequency filter on $|\mathscr{F}(O)|$\cite{Chen:2020,RN127}. It indicates that the object can be approximated with the speckle pattern $I$ and PhTF, $\Phi_{S}$:
\begin{equation}
\begin{aligned}
O\approx \mathscr{F}^{-1}(|\mathscr{F}(I)|\cdot e^{i\Phi_I-i\Phi_S}). \label{eq:OTF-P}
\end{aligned}
\end{equation}
As is shown in Fig. \ref{fig:MORE_block}, \emph{MORE} is proposed to generate the PhTF by iteratively computing the Fourier phase among the speckles. In this way, \emph{MORE} avoids the influence of the PSF autocorrelation introduced in traditional SCT\cite{RN31}, and achieves non-invasive imaging behind the scattering medium under the broadband illumination.

Similar as the Error-Reduction algorithm\cite{Bauschke:02}, the iterative process starts in the frequency domain with an initial random guess of PhTF, $\Phi_S$. In the k-th outer loop and j-th inner iterative loop, the Fourier transform of object, $\mathscr{F}(O_{k,j})$, is calculate with the frequency domain magnitude of speckle $|\mathscr{F}(I_j)|$ and difference of Fourier phase of speckle and PhTF, $\Phi_{I_j}-\Phi_{S_{k,j}}$. After applying an inverse Fourier transform of the object in the frequency domain, we use a real and nonnegative constraint of to update the guessed object, $O_{k,j}'$, in object domain. To return to the frequency domain, we performe a Fourier transform on $O_{k,j}'$ and calculate PhTF with speckle's phase of Fourier transform, $\Phi_{S_{k,j+1}}=\Phi_{I_j}-\Phi_{O_{k,j}'}$. Then the inner iterative move forward and turn to next frame until all the frames are used. After j-th inner iterative loop is completed, $k=k+1$ and another time outer loop starts and it goes from the first frame to the last one. When k reaches the number of loops we specify, we can obtain the PhTF of the imaging system, $\Phi_{S}$. The objects can be recovered with speckle frames and the PhTF, as Eq. (\ref{eq:OTF-P}) shows.

\begin{figure}[htbp]
\centering
\includegraphics[width=12cm]{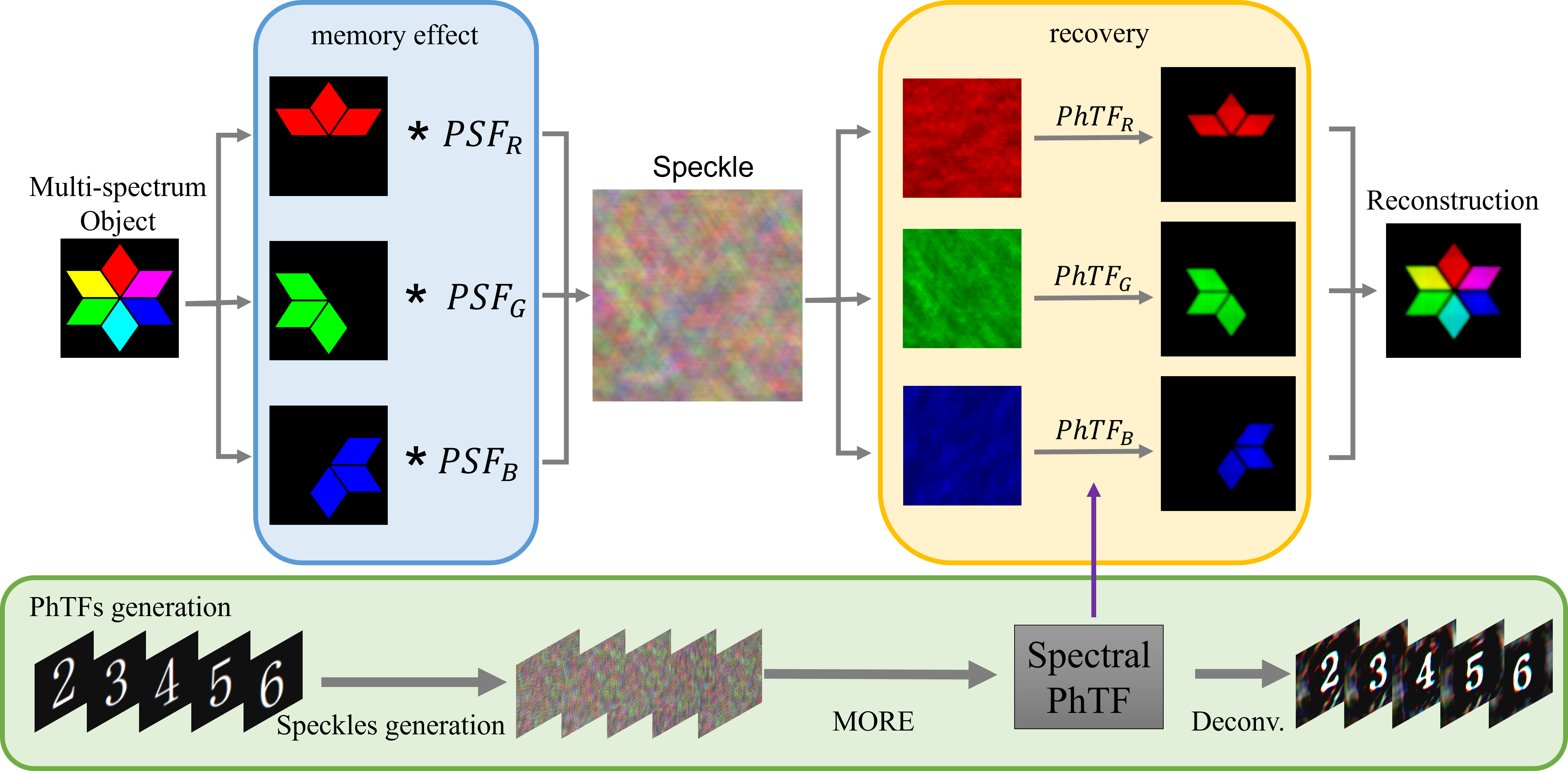}
\caption{\label{fig:simu}Numerical simulation of \emph{MORE} and color imaging by deconvolution with PhTF generated by \emph{MORE}. Color speckles are generated by convoluting with the different spectral PSF to simulate memory effect. Then we obtain the spectral PhTF by applying \emph{MORE} on the speckles. Object is recovered by deconvoluting with those PhTF retrieved by \emph{MORE}.}
\end{figure}

Additionally, by performing OTF retrieval on each of the three broadband spectrum channels separately, \emph{MORE} can be used for color imaging. Under broadband illumination, speckle pattern is the sum of speckles under narrow band illumination\cite{RN65,RN91}. For the speckle captured by a 3 channel R, G, B color camera, it can be regarded as a sum of the three broadband speckles, which can be described by Eq. (\ref{eq:spectral-speckle}). By applying \emph{MORE} in three broadband color channel (R, G, B), we can obtain different PhTF of those channels. Then the object can be retrieved by deonvoluting with PhTF. Equation. (\ref{eq:spectral-recovery}) shows the deconvolution operation in Fourier domain, where Fourier transform of object is equal to the Fourier transform of speckle divided with PhTF.

\begin{equation}
\begin{aligned}
I=\sum_{\lambda}^{R,G,B}I_{\lambda}=\sum_{\lambda}^{R,G,B}O_{\lambda} \ast S_{\lambda} \label{eq:spectral-speckle}
\end{aligned}
\end{equation}

\begin{equation}
\begin{aligned}
O=\sum_{\lambda}^{R,G,B}O_{\lambda}
=\sum_{\lambda}^{R,G,B}\mathscr{F}^{-1}(\mathscr{F}(O_{\lambda}))
=\sum_{\lambda}^{R,G,B} \mathscr{F}^{-1}(\frac{\mathscr{F}(I_{\lambda})}{e^{-i\Phi_{s_{\lambda}}}})\label{eq:spectral-recovery}
\end{aligned}
\end{equation}
Where $\lambda$ represents the broadband wavelength ranges of three color channels (R, G, B), $I$ is the color speckle with 3 channels captured by camera, $I_{\lambda}$ is the speckle intensity under wavelength range $\lambda$, $S_{\lambda}$ is the PSF of imaging system under wavelength range of $\lambda$, $O_{\lambda}$ is the object under wavelength range of $\lambda$, $\mathscr{F}$ and $\mathscr{F}^{-1}$ denotes the Fourier transform and inverse Fourier transform, respectively.

The recovery pipeline has been explicitly explained using a numerical simulation, as shown in Fig. \ref{fig:simu}. The object we demonstrate in simulation is a multicolored star. Its three color channels are extracted and combined using three separated PSFs generated randomly. Then, to get three single channel images, we performed a deconvolution operation using only the PhTF, which we have obtained on the R, G, and B images of speckles. By superposing those images, object's color image was rebuilt. 

\section{Experimental results}
\label{Experimental Results and Discussion}
\subsection{Amplitude object imaging under LED illumination}

\begin{figure}[htbp]
\centering
\includegraphics[width=12cm]{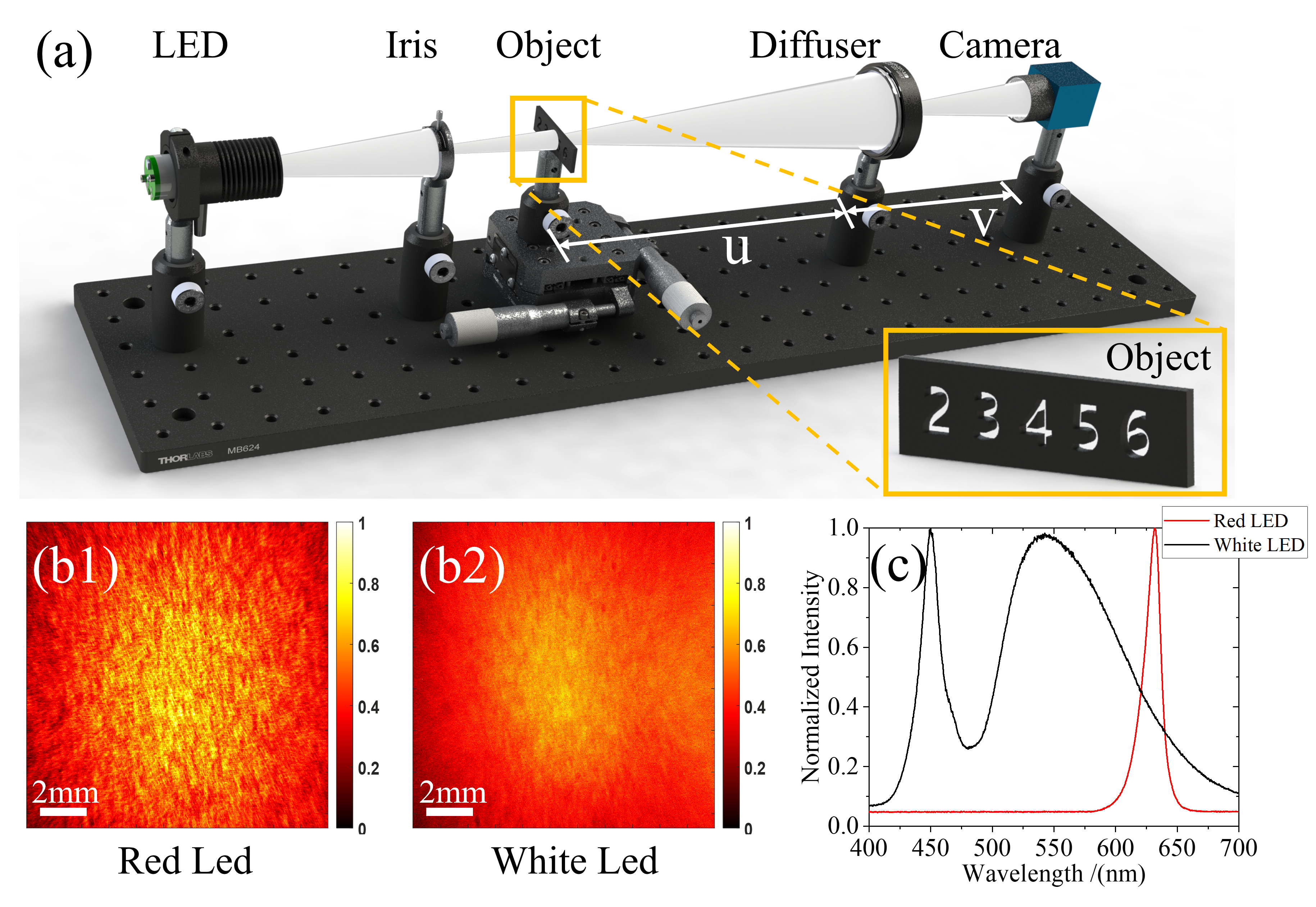}
\caption{\label{fig:experiment}Experiment setup of dynamic imaging under white illumination. (a) Experiment setup schematic of amplitude object imaging under different bandwidth LED. (b)Raw speckles of amplitude number object '2' captured by monochromatic camera under 500 ms exposure time, (b1) under red light LED (b2) under white light LED. (c) The spectrum of white and red light LED measured in our experiments.}
\end{figure}

With the setup shown in Fig. \ref{fig:experiment}, we demonstrate non-invasive imaging through a scattering medium under white-light illumination with multi-frame speckle. A white and a red light LED were selected as the light sources. LED's full width at half maximum of spectrum are about 200 nm and 15 nm, as shown in Fig. \ref{fig:experiment}(c). Several number objects (2.4-mm wide and 3.6-mm high numbers, Fig. \ref{fig:experiment}(a)) were placed at a distance $u$ = 200 mm respectively from the diffuser. A CMOS camera (5496 $\times$ 3672 px with a pixel size of 2.4 $\times$ 2.4 $\mu$m) was placed $v$ = 100 mm respectively from the diffuser. Diffuser is a ground glass of 2-mm thickness and 220 grit. The magnification of scattering imaging system was $M=v/u=100 \, mm / 200\, mm = 0.5$.

As the light carrying the object information passes through the scattering medium, a specific pattern associated with the object was formed, which is captured by the camera behind the scattering medium. After collecting a series of speckles of objects (different number objects '2', '3', '4', '5' and '6' replaced in our experiment), the raw camera image, Fig. \ref{fig:experiment}(b), was spatial normalized by dividing the raw camera image by a low-pass filter version of it. And then the speckle patterns are smoothed by a Gaussian kernel filter (size: 20 pixels) with a standard-deviation width of 2 pixels to filter out high frequency noise. Note that the support mat constraint is important to the phase retrieval process. The closer the support mat is to the direct imaging size, the more accurate the recovered image will be\cite{Bauschke:02}. Even the support mat ($200\times 200$ pixels) is set to be twice object true size ($80\times120$ pixels),  \emph{MORE} can still provides a reasonable result.

\begin{figure}[htbp]
\centering
\includegraphics[width=12cm]{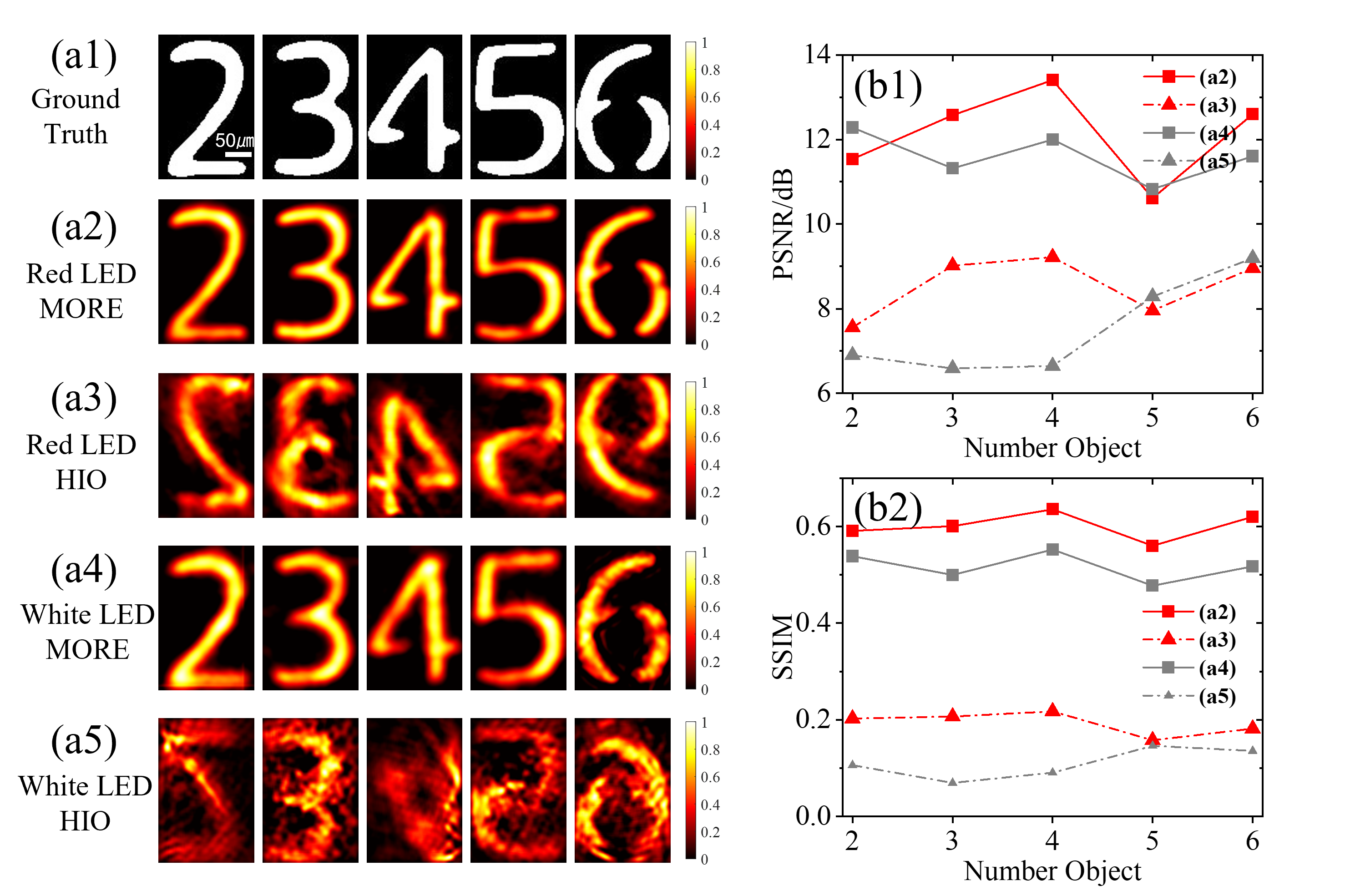}
\caption{\label{fig:MORE_PR}Comparison between \emph{MORE} and SCT with phase retrieval algorithm (HIO) under an exposure time of 500 ms. (a1) the ground truth images of '2','3','4','5','6'. (a2) Illuminated by a red light LED (wavelength: 600-650 nm), the recovery images from 5 frames and 5 iterative loops of \emph{MORE} (equal to 25 times of iterations), scale bar: 50 \textmu m. (a3) The best recovered images of the same data collected in (a2) by HIO phase retrieval algorithm with 50000 times iteration. (a4) Under the illumination of a white LED (wavelength: 400-700 nm), the reconstruction of \emph{MORE} with 5 frames and 5 iterative loops. (a5) Results recovered by HIO algorithm. (b) PSNR and SSIM of the images in (a2-a5). All the images used in image quality analysis are rotated to the same orientation as ground truth.} 
\end{figure}

The recovered results of \emph{MORE} and SCT with phase retrieval algorithm(HIO, hybrid-input-output algorithm) are shown in Fig. \ref{fig:MORE_PR}(a). Under the illumination of 15 nm and 200 nm, results obtained by \emph{MORE} are better than those got from SCT with phase retrieval algorithm (HIO). The results of HIO shown in Figs. \ref{fig:MORE_PR}(a3) and (a5) are the best recovery images in several trials. There are 50000 iterations for each trail. By comparing \emph{MORE}'s recovery of the numbers with the objects' ground truth, the object can be correctly identified. It is worth noting that results from \emph{MORE} are not selected by us manually and all the recovered images having the correct orientation which is missing in the SCT with phase retrieval process\cite{Fienup:82}. From Fig. \ref{fig:MORE_PR}(b), both \emph{MORE} and HIO achieve a better quality under narrower band illumination. As the spectrum becomes wider, the contrast of speckle would be lower, which causes instability of speckle correlation and phase retrieval. However, \emph{MORE} can still work well under both narrow and broadband illumination. Additionally, we calculate the peak-signal-to-noise ratio (PSNR) and structural similarity index measure (SSIM)\cite{ImageQuality} of reconstructed image to evaluate the quality of results. And On both PSNR and SSIM, the image of recovered by \emph{MORE} is better than by HIO among all the five numbers. It shows that our method does work better than SCT and phase retrieval algorithm under broadband illumination.

\subsection{Multispectral object imaging}

\begin{figure}[htbp]
\centering
\includegraphics[width=12cm]{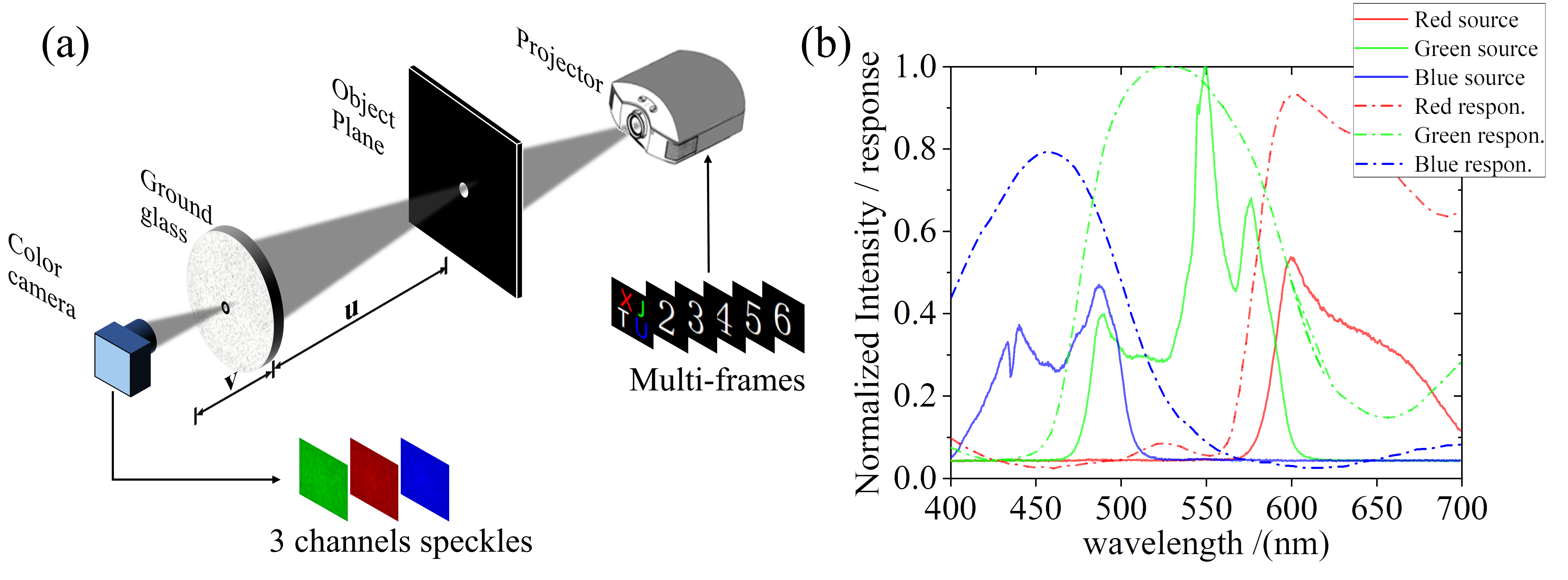}
\caption{\label{fig:color_imaging}(a) Experiment setup of color imaging. (b) Source spectrum and camera response to spectrum. Solid lines are the spectrum of projector source, which are measured by projecting monochromatic red blue and green images to the fiber spectrometer.}
\end{figure}

Furthermore, we perform an non-invasive broadband color imaging experiments with \emph{MORE}. The experiment scheme is shown in Fig. \ref{fig:color_imaging}(a). The light source and object are replaced by a projector with three broad-spectrum LED light sources. We utilized it to project a series of number objects similar to those in Fig.\ref{fig:experiment}. Then we took a serials of pictures of the speckle patterns behind the ground glass with a color camera.

\begin{figure}[htbp]
\centering
\includegraphics[width=12cm]{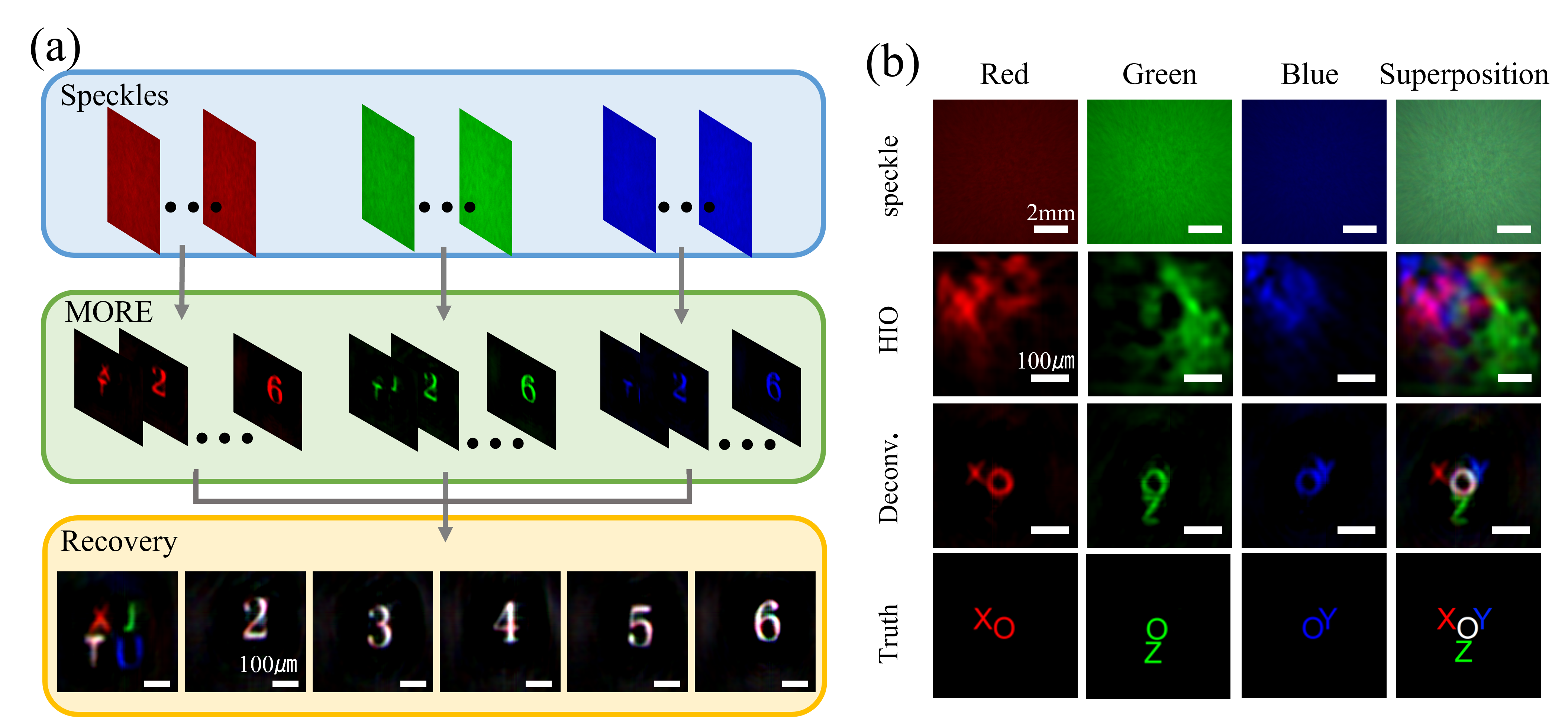}
\caption{\label{fig:color_recovery}Result of color imaging and the measurements of spectrum. (a) Reconstruction pipeline of color imaging. Objects are recovered by superpositing 3 channels image retrieved by \emph{MORE}, Scale bar: 100 \textmu m. (b) Speckle of single frame of color object 'XYOZ', Scale bar: 2 mm.The result of single-shot HIO and the recovery deconvolved with PhTF which generated by \emph{MORE}, Scale bar: 100 \textmu m.} 
\end{figure}

As Fig. \ref{fig:color_recovery}(a) shows, we extract R, G, and B color channels from the color speckle and obtain PhTF of broadband spectrum by repeating the process mentioned above. Additionally, from the spectrum of projector and camera spectral response curve in Fig. \ref{fig:color_recovery}(b), blue and red spectrum intensity is comparatively lower than green spectrum intensity when projector brightness is the same. As a result, the gain of the various channels need be adjusted manually to apply the white balance to color camera. We conduct a deconvolution operation with the PhTF produced previously to recover three monochromatic R, G, and B images. Then the color images of white numbers '2','3,'4','5','6' and colorful text 'XJTU' are superimposed by three channels images recovered from three broadband spectrum speckles.

After we obtain the PhTF of the non-invasion scattering imaging system, we conduct another experiment which use another multi-spectral object to evaluate the correctness of PhTF generated by \emph{MORE}. In Fig. \ref{fig:color_recovery}(b), three channels of speckle of colorful text 'XYOZ' are deconvoluted with spectral PhTF obtained by speckles from \ref{fig:color_recovery}(a). Compared with the truth of object, the images of channels are recovered successfully by deconvoluting with PhTF generated by \emph{MORE}, while results obtained with phase retrieval algorithm (HIO) fails to recovery the object.

\section{Discussion}
\begin{figure}[htbp]
\centering
\includegraphics[width=12cm]{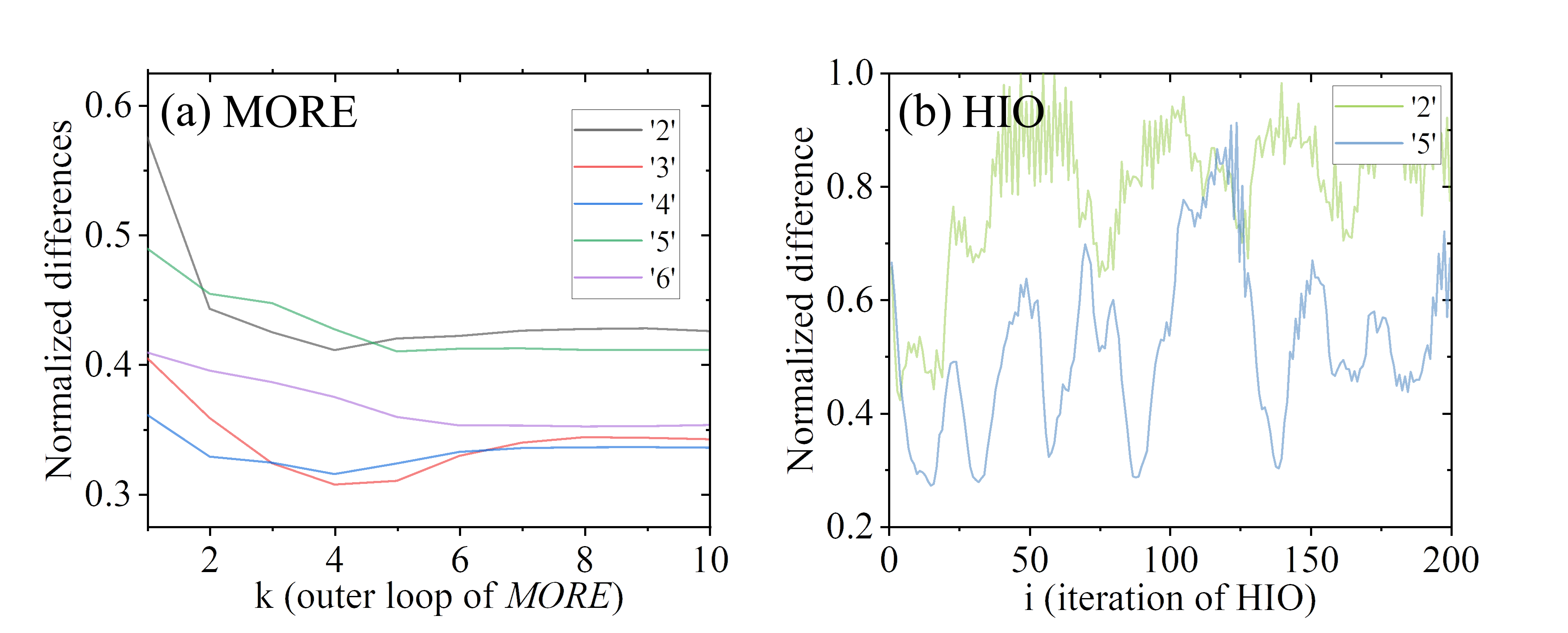}
\caption{\label{fig:errF}Normalized difference curves of the reconstruction of number objects of \emph{MORE} and HIO. Data are taken with 500ms exposure time under white illumination. (a)The convergence curves of the recovery for '2' to '6' in \emph{MORE}. 'k' represents times of the outer loop in \emph{MORE} and each outer loop contains 5 times iteration operations. (b) Normalized error the of '2' and '5' in HIO. 'i' is the total iteration times in HIO.}
\end{figure}

The broadband imaging method we proposed is based on the convolution operation between the object and the PSF within memory effect, as Eq. (\ref{eq:fourier(I=O*S)}) shows\cite{RN125}. Compared to the SCT with phase retrieval algorithm, \emph{MORE} is more like a way to measure PhTF by a serials of frames or dynamic object. Non-invasive color imaging  can be achieved by using narrowband light source or adding different narrowband filters before camera\cite{RN115}. However there is no demand of narrowband source or filters in our method. It has been proved that by modifying phase retrieval algorithm with PhTF, object can be successfully recovered with single frame\cite{RN91}. However \emph{MORE} has a faster convergence speed in iteration process and provides more stable and reliable results than modified phase retrieval algorithm. It is because that there is more redundant information of PhTF contained in multi-frame speckles than single-shot frame. As shown in Fig. \ref{fig:errF}, \emph{MORE} converges from the first outer iterative loop, and  stably reaches its global minimum after the fifth outer loop (equals to 25 times iteration). The SCT with phase retrieval algorithm (HIO) remain unstable even after the 200 times iteration.

 A projector has been used as the multi-spectral object and the incoherent light source to simply demonstrate \emph{MORE} in our work. However, \emph{MORE} we suggested can still work, when a small colored object within the memory effect is illuminated with white light LED or even natural light. It is important to note that one key benefit of the deconvolution scattering imaging method is the advantage to determine an object's relative position and orientation. So the spatial position relationship of those items within the memory effect can also be obtained using \emph{MORE}. Beacuse all the channels of color speckles and spectral PhTF are calculated separately, it is inevitable to superposing color image manually. 
 
\section{Conclusion}
In conclusion, we prove that our multi-frame OTF retrieve engine (\emph{MORE}) can be used in non-invasive mono and color imaging under broadband illumination successfully. Compared with existing methods\cite{RN91,RN40,RN98,RN65}, it has no need for prior information of PSF and it achieves a faster and stable convergence and reliable results due to redundant information. \emph{MORE} overcomes the strict requirements of narrow band illumination and long time consumption on recovery process. It is a effective way to achieve color scattering imaging under broadband illumination.


\begin{thebibliography}{10}
\newcommand{\enquote}[1]{``#1''}

\bibitem{Speckle:Phenomena}
J.~W. Goodman, \emph{Speckle Phenomena in Optics: Theory and Applications}
  (SPIE Press, 2020), 2nd ed.

\bibitem{RN35}
D.~Faccio, A.~Velten, and G.~Wetzstein, \enquote{Non-line-of-sight imaging,}
  {\protect\JournalTitle{Nature Reviews Physics}} \textbf{2}, 318--327 (2020).

\bibitem{RN85}
T.~Wu, J.~Dong, and S.~Gigan, \enquote{Non-invasive single-shot recovery of a
  point-spread function of a memory effect based scattering imaging system,}
  {\protect\JournalTitle{Opt Lett}} \textbf{45}, 5397--5400 (2020). Wu, Tengfei
  Dong, Jonathan Gigan, Sylvain eng Opt Lett. 2020 Oct 1;45(19):5397-5400. doi:
  10.1364/OL.400869.

\bibitem{RN89}
G.~Barbastathis, A.~Ozcan, and G.~Situ, \enquote{On the use of deep learning
  for computational imaging,} {\protect\JournalTitle{Optica}} \textbf{6}
  (2019).

\bibitem{RN51}
O.~Katz, E.~Small, and Y.~Silberberg, \enquote{Looking around corners and
  through thin turbid layers in real time with scattered incoherent light,}
  {\protect\JournalTitle{Nature Photonics}} \textbf{6}, 549--553 (2012).

\bibitem{RN52}
I.~M. Vellekoop and A.~P. Mosk, \enquote{Focusing coherent light through opaque
  strongly scattering media,} {\protect\JournalTitle{Opt. Lett.}} \textbf{32},
  2309--2311 (2007).

\bibitem{RN122}
A.~P. Mosk, A.~Lagendijk, G.~Lerosey, and M.~Fink, \enquote{Controlling waves
  in space and time for imaging and focusing in complex media,}
  {\protect\JournalTitle{Nature Photonics}} \textbf{6}, 283--292 (2012).

\bibitem{RN123}
S.~M. Popoff, G.~Lerosey, R.~Carminati, M.~Fink, A.~C. Boccara, and S.~Gigan,
  \enquote{Measuring the transmission matrix in optics: an approach to the
  study and control of light propagation in disordered media,}
  {\protect\JournalTitle{Phys Rev Lett}} \textbf{104}, 100601 (2010).

\bibitem{RN124}
D.~Andreoli, G.~Volpe, S.~Popoff, O.~Katz, S.~Grésillon, and S.~Gigan,
  \enquote{Deterministic control of broadband light through a multiply
  scattering medium via the multispectral transmission matrix,}
  {\protect\JournalTitle{Scientific Reports}} \textbf{5}, 10347 (2015).

\bibitem{RN65}
S.~K. Sahoo, D.~Tang, and C.~Dang, \enquote{Single-shot multispectral imaging
  with a monochromatic camera,} {\protect\JournalTitle{Optica}} \textbf{4}
  (2017).

\bibitem{RN98}
H.~Zhuang, H.~He, X.~Xie, and J.~Zhou, \enquote{High speed color imaging
  through scattering media with a large field of view,}
  {\protect\JournalTitle{Sci Rep}} \textbf{6}, 32696 (2016).

\bibitem{RN31}
O.~Katz, P.~Heidmann, M.~Fink, and S.~Gigan, \enquote{Non-invasive single-shot
  imaging through scattering layers and around corners via speckle
  correlations,} {\protect\JournalTitle{Nature Photonics}} \textbf{8}, 784--790
  (2014).

\bibitem{Li:19}
X.~Li, J.~A. Greenberg, and M.~E. Gehm, \enquote{Single-shot multispectral
  imaging through a thin scatterer,} {\protect\JournalTitle{Optica}}
  \textbf{6}, 864--871 (2019).

\bibitem{RN91}
D.~Lu, Q.~Xing, M.~Liao, G.~Situ, X.~Peng, and W.~He, \enquote{Single-shot
  noninvasive imaging through scattering medium under white-light
  illumination,} {\protect\JournalTitle{Opt Lett}} \textbf{47}, 1754--1757
  (2022).

\bibitem{RN40}
S.~Zheng, M.~Liao, F.~Wang, W.~He, X.~Peng, and G.~Situ,
  \enquote{Non-line-of-sight imaging under white-light illumination: a two-step
  deep learning approach,} {\protect\JournalTitle{Opt Express}} \textbf{29},
  40091--40105 (2021).

\bibitem{RN101}
E.~Guo, Y.~Sun, S.~Zhu, D.~Zheng, C.~Zuo, L.~Bai, and J.~Han,
  \enquote{Single-shot color object reconstruction through scattering medium
  based on neural network,} {\protect\JournalTitle{Optics and Lasers in
  Engineering}} \textbf{136} (2021).

\bibitem{Fienup:82}
J.~R. Fienup, \enquote{Phase retrieval algorithms: a comparison,}
  {\protect\JournalTitle{Appl. Opt.}} \textbf{21}, 2758--2769 (1982).

\bibitem{MORRIS1997197}
G.~A. Morris, H.~Barjat, and T.~J. Home, \enquote{Reference deconvolution
  methods,} {\protect\JournalTitle{Progress in Nuclear Magnetic Resonance
  Spectroscopy}} \textbf{31}, 197--257 (1997).

\bibitem{RN106}
L.~Zhu, Y.~Wu, J.~Liu, T.~Wu, L.~Liu, and X.~Shao, \enquote{Color imaging
  through scattering media based on phase retrieval with triple correlation,}
  {\protect\JournalTitle{Optics and Lasers in Engineering}} \textbf{124}
  (2020).

\bibitem{RN37}
T.~Wu, O.~Katz, X.~Shao, and S.~Gigan, \enquote{Single-shot diffraction-limited
  imaging through scattering layers via bispectrum analysis,}
  {\protect\JournalTitle{Opt Lett}} \textbf{41}, 5003--5006 (2016).

\bibitem{Chen:2020}
Y.~Yuan and H.~Chen, \enquote{Dynamic noninvasive imaging through turbid media
  under low signalnoise-ratio,} {\protect\JournalTitle{New J. Phys}}
  \textbf{22}, 093046 (2020).

\bibitem{RN127}
S.~Mukherjee, A.~Vijayakumar, M.~Kumar, and J.~Rosen, \enquote{3d imaging
  through scatterers with interferenceless optical system,}
  {\protect\JournalTitle{Sci Rep}} \textbf{8}, 1134 (2018).

\bibitem{Bauschke:02}
H.~H. Bauschke, P.~L. Combettes, and D.~R. Luke, \enquote{Phase retrieval,
  error reduction algorithm, and fienup variants: a view from convex
  optimization,} {\protect\JournalTitle{J. Opt. Soc. Am. A}} \textbf{19},
  1334--1345 (2002).

\bibitem{ImageQuality}
A.~Horé and D.~Ziou, \enquote{Image quality metrics: Psnr vs. ssim,} in
  \emph{2010 20th International Conference on Pattern Recognition,}  (2010),
  pp. 2366--2369.

\bibitem{RN125}
D.~Wang, S.~K. Sahoo, X.~Zhu, G.~Adamo, and C.~Dang, \enquote{Non-invasive
  super-resolution imaging through dynamic scattering media,}
  {\protect\JournalTitle{Nature Communications}} \textbf{12}, 3150 (2021).

\bibitem{RN115}
L.~Zhu, Y.~Wu, J.~Liu, T.~Wu, L.~Liu, and X.~Shao, \enquote{Color imaging
  through scattering media based on phase retrieval with triple correlation,}
  {\protect\JournalTitle{Optics and Lasers in Engineering}} \textbf{124}
  (2020).

\end{thebibliography}

\end{document}